# COMPLEXITY OF SELF-ASSEMBLED SHAPES[*]

DAVID SOLOVEICHIK[†] AND ERIK WINFREE[‡]

**Abstract.** The connection between self-assembly and computation suggests that a shape can be considered the output of a self-assembly "program," a set of tiles that fit together to create a shape. It seems plausible that the size of the smallest self-assembly program that builds a shape and the shape's descriptional (Kolmogorov) complexity should be related. We show that when using a notion of a shape that is independent of scale, this is indeed so: in the Tile Assembly Model, the minimal number of distinct tile types necessary to self-assemble a shape, at some scale, can be bounded both above and below in terms of the shape's Kolmogorov complexity. As part of the proof of the main result, we sketch a general method for converting a program outputting a shape as a list of locations into a set of tile types that self-assembles into a scaled up version of that shape. Our result implies, somewhat counter-intuitively, that self-assembly of a scaled-up version of a shape often requires fewer tile types. Furthermore, the independence of scale in self-assembly theory appears to play the same crucial role as the independence of running time in the theory of computability. This leads to an elegant formulation of languages of shapes generated by self-assembly. Considering functions from integers to shapes, we show that the running-time complexity, with respect to Turing machines, is polynomially equivalent to the scale complexity of the same function implemented via self-assembly by a finite set of tile types. Our results also hold for shapes defined by Wang tiling – where there is no sense of a self-assembly process – except that here time complexity must be measured with respect to non-deterministic Turing machines.

**Key words.** Kolmogorov complexity, scaled shapes, self-assembly, Wang tiles

**AMS subject classifications.** 68Q30, 68Q05, 52C20, 52C45

**1. Introduction.** Self-assembly is the process by which an organized structure can spontaneously form from simple parts. The Tile Assembly Model [21, 20], based on Wang tiling [19], formalizes the two-dimensional self-assembly of square units called "tiles" using a physically plausible abstraction of crystal growth. In this model, a new tile can adsorb to a growing complex if it binds strongly enough. Each of the four sides of a tile has an associated bond type that interacts with a certain strength with matching sides of other tiles. The process of self-assembly is initiated by a single seed tile and proceeds via the sequential addition of new tiles. Confirming the physical plausibility and relevance of the abstraction, simple self-assembling systems of tiles have been built out of certain types of DNA molecules [22, 15, 14, 12, 10]. The possibility of using self-assembly for nanofabrication of complex components such as circuits has been suggested as a promising application [6].

The view that the "shape" of a self-assembled complex can be considered the output of a computational process [2] has inspired recent interest [11, 1, 3, 9, 4]. While it was shown through specific examples that self-assembly can be used to construct interesting shapes and patterns, it was not known in general which shapes could be self-assembled from a small number of tile types. Understanding the complexity of shapes is facilitated by an appropriate definition of shape. In our model, a tile system generates a particular shape if it produces any scaled version of that shape (§3). This definition may be thought to formalize the idea that a structure can be made up of arbitrarily small pieces. Computationally, it is analogous to disregarding computation time and is thus more appropriate as a notion of output of a *universal* computation process.[1] Using this definition of shape, we show that for any shape $\tilde{S}$, if $K_{sa}(\tilde{S})$ is the minimal number of distinct tile types necessary to self-assemble it, then $K_{sa}(\tilde{S}) \log K_{sa}(\tilde{S})$ is within multiplicative and additive constants (independent of $\tilde{S}$) of the shape's Kolmogorov complexity. This theorem is proved by construction (which might be of independent interest) of a general method for converting a program that outputs a fixed size shape as a list of locations into a tile system that self-assembles a scaled version of the shape. Our result ties the computation of a shape and its self-assembly, and, somewhat counter-intuitively, implies that it may often require fewer tile types to self-assemble a larger instance of a shape than a smaller instance thereof. Another consequence of the theorem is that the minimal number of tile types necessary to self-assemble an arbitrary scaling of a shape is uncomputable. Answering the same question about shapes of a fixed size is computable but NP complete [1].

---

[*]The extended abstract version of this paper was published in the Proceedings of DNA Computing 10. This work was supported by NSF CAREER Grant No. 0093486.
[†]California Institute of Technology, Pasadena, CA 91125, USA `dsolov@caltech.edu`
[‡]California Institute of Technology, Pasadena, CA 91125, USA `winfree@caltech.edu`

**2. The Tile Assembly Model.** We present a description of the Tile Assembly Model based on Rothemund and Winfree [11] and Rothemund [9]. We will be working on a $\mathbb{Z} \times \mathbb{Z}$ grid of unit square locations. The **directions** $\mathcal{D} = \{N, E, W, S\}$ are used to indicate relative positions in the grid. Formally, they are functions $\mathbb{Z} \times \mathbb{Z} \to \mathbb{Z} \times \mathbb{Z}$: $N(i,j) = (i, j+1)$, $E(i,j) = (i+1, j)$, $S(i,j) = (i, j-1)$, and $W(i,j) = (i-1, j)$. The inverse directions are defined naturally: $N^{-1}(i,j) = S(i,j)$, etc. Let $\Sigma$ be a set of **bond types**. A **tile type** $\bar{t}$ is a 4-tuple $(\sigma_N, \sigma_E, \sigma_S, \sigma_W) \in \Sigma^4$ indicating the associated bond types on the north, east, south, and west sides. Note that tile types are oriented, so a rotated version of a tile type is considered to be a different tile type. A special bond type $null$ represents the lack of an interaction and the special tile type $empty = (null, null, null, null)$ represents an empty space. If $T$ is a set of tile types, a **tile** is a pair $(\bar{t}, (i,j)) \in T \times \mathbb{Z}^2$ indicating that location $(i,j)$ contains the tile type $\bar{t}$. Given the tile $t = (\bar{t}, (i,j))$, $type(t) = \bar{t}$ and $pos(t) = (i,j)$. Further, $bond_D(\bar{t})$, where $D \in \mathcal{D}$, is the bond type of the respective side of $\bar{t}$, and $bond_D(t) = bond_D(type(t))$. A **configuration** is a set of non-$empty$ tiles, with types from $T$, such that there is no more than one tile in every location $(i,j) \in \mathbb{Z} \times \mathbb{Z}$. For any configuration $A$, we write $A(i,j)$ to indicate the tile at location $(i,j)$ or the tile $(empty, (i,j))$ if there is no tile in $A$ at this location.

A **strength function** $g: \Sigma \times \Sigma \to \mathbb{Z}$, where $null \in \Sigma$, defines the interactions between adjacent tiles: we say that a tile $t_1$ interacts with its neighbor $t_2$ with strength $\Gamma(t_1, t_2) = g(\sigma, \sigma')$ where $\sigma$ is the bond type of tile $t_1$ that is adjacent to the bond type $\sigma'$ of tile $t_2$.[2] The $null$ bond has a zero interaction strength (i.e. $\forall \sigma \in \Sigma$, $g(null, \sigma) = 0$). We say that a strength function is **diagonal** if it is non-zero only for $g(\sigma, \sigma')$ such that $\sigma = \sigma'$. Unless otherwise noted, a tile system is assumed to have a diagonal strength function. Our constructions use diagonal strength functions with the range $\{0, 1, 2\}$. We say that a bond type $\sigma$ has **strength** $g(\sigma, \sigma)$. Two tiles are **bonded** if they interact with a positive strength. For a configuration $A$, we use the notation $\Gamma_D^A(t) = \Gamma(t, A(D(pos(t))))$.[3] For $L \subseteq \mathcal{D}$ we define $\Gamma_L^A(t) = \sum_{D \in L} \Gamma_D^A(t)$.

---

[1]The production of a shape of a fixed size cannot be considered the output of a universal computation process. Whether a universal process will output a given shape is an undecidable question, whereas this can be determined by exhaustive enumeration in the Tile Assembly Model.

[2]More formally,

$$\Gamma(t_1, t_2) = \begin{cases} g(bond_{D^{-1}}(t_1), bond_D(t_2)) & \text{if } \exists D \in \mathcal{D} \text{ s.t. } pos(t_1) = D(pos(t_2)); \\ 0 & \text{otherwise.} \end{cases}$$

[3]Note that $t \neq A(pos(t))$ is a valid choice. In that case $\Gamma_D^A(t)$ tells us how $t$ would bind if it were in A.

A **tile system** **T** is a quadruple $(T, t_s, g, \tau)$ where $T$ is a finite set of non-empty tile types, $t_s$ is a special **seed tile** with $type(t_s) \in T$, $g$ is a strength function, and $\tau$ is the threshold parameter. Self-assembly is defined by a relation between configurations. Suppose $A$ and $B$ are two configurations, and $t$ is a tile such that $A = B$ except at $pos(t)$ and $A(pos(t)) = null$ but $B(pos(t)) = t$. Then we write $A \to_\mathbf{T} B$ if $\Gamma^A_\mathcal{D}(t) \geq \tau$. This means that a tile can be added to a configuration iff the sum of its interaction strengths with its neighbors reaches or exceeds $\tau$. The relation $\to^*_\mathbf{T}$ is the reflexive transitive closure of $\to_\mathbf{T}$.

Whereas a configuration can be any arrangement of tiles (not necessarily connected), we are interested in the subclass of configurations that can result from a self-assembly process. Formally, the tile system and the relation $\to^*_\mathbf{T}$ define the partially ordered set of **assemblies**: $Prod(\mathbf{T}) = \{A \text{ s.t. } \{t_s\} \to^*_\mathbf{T} A\}$, and the set of **terminal assemblies**: $Term(\mathbf{T}) = \{A \in Prod(\mathbf{T}) \text{ and } \nexists B \neq A \text{ s.t. } A \to^*_\mathbf{T} B\}$. A tile system **T** **uniquely produces** $A$ if $\forall B \in Prod(\mathbf{T}), B \to^*_\mathbf{T} A$ (which implies $Term(\mathbf{T}) = \{A\}$).

An **assembly sequence** $\vec{A}$ of **T** is a sequence of pairs $(A_n, t_n)$ where $A_0 = \{t_0\} = \{t_s\}$ and $A_{n-1} \to_\mathbf{T} A_n = A_{n-1} \cup \{t_n\}$. Here we will exclusively consider finite assembly sequences. If a finite assembly sequence $\vec{A}$ is implicit, $A$ indicates the last assembly in the sequence.

The tile systems used in our constructions have $\tau = 2$ with the strength function ranging over $\{0, 1, 2\}$. It is known that $\tau = 1$ systems with strength function ranging over $\{0, 1\}$ are rather limited [11, 9]. In our drawings, the bond type $\sigma$ may be illustrated by a combination of shading, various graphics and symbols. Strength-2 bond type will always contain two dots in their representation. All markings must match for two bond types to be considered identical. For example, the north bond type of the following tile has strength 2 and the others have strength 1.

The constructions in this paper do not use strength-0 bond types (other than in *empty* tiles); thus, there is no confusion between strength-1 and strength-0 bond types. Zero strength interactions due to mismatches between adjacent tiles do occur in our constructions.

**2.1. Guaranteeing Unique Production.** When describing tile systems that produce a desired assembly, we would like an easy method for showing that this assembly is uniquely produced. While it might be easy to find an assembly sequence that leads to a particular assembly, there might be many other assembly sequences that lead elsewhere. Here we present a property of an assembly sequence that guarantees that the assembly it produces is indeed the uniquely produced assembly of the tile system.

Rothemund [9] describes the deterministic-RC property of an assembly that guarantees its unique production and that is very easy to check; however, this property is satisfied only by assemblies that have no strength-0 interactions between neighboring non-empty tiles and is not directly applicable to the assemblies we will consider here. A more general poly-time test for unique production was also shown by Rothemund [9], but it can be difficult to prove that a particular assembly would satisfy this test. On the other hand, the notion of locally deterministic assembly sequences introduced here is easily checkable and sufficient for the constructions in this paper.

DEFINITION 2.1. *For an assembly sequence $\vec{A}$ we define the following sets of directions for $\forall i, j \in \mathbb{Z}$, letting $t = A(i, j)$:*

- $\mathit{inputsides}^{\vec{A}}(t) = \{D \in \mathcal{D} \text{ s.t. } t = t_n \text{ and } \Gamma^{A_n}_D(t_n) > 0\}$,
- $\mathit{propsides}^{\vec{A}}(t) = \{D \in \mathcal{D} \text{ s.t. } D^{-1} \in \mathit{inputsides}^{\vec{A}}(A(D(pos(t))))\}$,
- $\mathit{termsides}^{\vec{A}}(t) = \mathcal{D} - \mathit{inputsides}^{\vec{A}}(t) - \mathit{propsides}^{\vec{A}}(t)$.

Intuitively, *inputsides* are the sides with which the tile initially binds in the process of self-assembly; these sides determine its identity. *propsides* propagate information by being the sides to which neighboring tiles bind. *termsides* are sides that do neither. Note that by definition *empty* tiles have four *termsides*.

DEFINITION 2.2. *A finite assembly sequence $\vec{A}$ of $\mathbf{T} = (T, t_s, g, \tau)$ is called **locally deterministic** if $\forall i, j \in \mathbb{Z}$, letting $t = A(i, j)$,*

1. $\Gamma^{A}_{\mathit{inputsides}^{\vec{A}}(t)}(t) \leq \tau$
2. $\forall t' \text{ s.t. } type(t') \in T, \, pos(t') = pos(t) \text{ but } type(t') \neq type(t)$,

$$\Gamma^{A}_{\mathcal{D} - \mathit{propsides}^{\vec{A}}(t)}(t') < \tau.$$

We allow the possibility of $<$ in property (1) in order to account for the seed and *empty* tiles. Intuitively, the first property says that when a new tile binds a growing assembly, it binds "just barely." The second property says that nothing can grow from non-propagating sides except "as desired." We say that $\mathbf{T}$ is locally deterministic if there exists a locally deterministic assembly sequence for it.

It is clear that if $\vec{A}$ is a locally deterministic assembly sequence of $\mathbf{T}$, then $A \in Term(\mathbf{T})$. Otherwise, the *empty* tile in the position where a new (non-empty) tile can be added to $A$ would violate the second property. However, the existence of a locally deterministic assembly sequence leads to a much stronger conclusion:

THEOREM 2.3. *If there exists a locally deterministic assembly sequence $\vec{A}$ of $\mathbf{T}$ then $\mathbf{T}$ uniquely produces $A$.*

*Proof.* See Appendix A. □

**3. Arbitrarily Scaled Shapes and Their Complexity.** In this section, we introduce the model for the output of the self-assembly process used in this paper. Let $S$ be a finite set of locations on $\mathbb{Z} \times \mathbb{Z}$. The adjacency graph $G(S)$ is the graph on $S$ defined by the adjacency relation where two locations are considered adjacent if they are directly north/south, or east/west of one another. We say that $S$ is a **coordinated shape** if $G(S)$ is connected.[4] The **coordinated shape of assembly** $A$ is the set $S_A = \{pos(t) \text{ s.t. } t \in A\}$. Note that $S_A$ is a coordinated shape because $A$ contains a single connected component.

For any set of locations $S$, and any $c \in \mathbb{Z}^+$, we define a $c$-**scaling of S** as

$$S^c = \{(i, j) \text{ s.t. } (\lfloor i/c \rfloor, \lfloor j/c \rfloor) \in S\}.$$

Geometrically, this represents a "magnification" of $S$ by a factor $c$. Note that a scaling of a coordinated shape is itself a coordinated shape: every node of $G(S)$ gets mapped to a $c^2$-node connected subgraph of $G(S^c)$ and the relative connectivity of the subgraphs is the same as the connectivity of the nodes of $G(S)$. A parallel argument shows that if $S^c$ is a coordinated shape, then so is $S$. We say that coordinated shapes $S_1$ and $S_2$ are **scale-equivalent** if $S_1^c = S_2^d$ for some $c, d \in \mathbb{Z}^+$. Two coordinated shapes are **translation-equivalent** if they can be made identical by translation. We write $S_1 \cong S_2$ if $S_1^c$ is translation-equivalent to $S_2^d$ for some $c, d \in \mathbb{Z}^+$. Scale-equivalence, translation-equivalence and $\cong$ are equivalence relations (Appendix B). This defines the equivalence classes of coordinated shapes under $\cong$. The equivalence class containing $S$ is denoted $\tilde{S}$ and we refer to it as the **shape** $\tilde{S}$. We say that $\tilde{S}$ is the **shape of assembly** $A$ if $S_A \in \tilde{S}$. The view of computation performed by the self-assembly process espoused here is the production of a shape as the "output" of the self-assembly process, with the understanding that the scale of the shape is irrelevant. Intuitively, this view is appropriate to the extent that a physical object can be constructed from arbitrarily small pieces.

---

[4] We say "coordinated" to make explicit that a fixed coordinate system is used. We reserve the unqualified term "shape" for when we ignore scale and translation.

Having defined the notion of shapes, we turn to their descriptional complexity. As usual, the Kolmogorov complexity of a binary string $x$ with respect to a universal Turing machine $U$ is $K_U(x) = \min\{|p| \text{ s.t. } U(p) = x\}$. (See the exposition of Li and Vitanyi [13] for an in-depth discussion of Kolmogorov complexity.) Let us fix some "standard" universal machine $U$. We call the Kolmogorov complexity of a coordinated shape $S$ to be the size of the smallest program outputting it as a list of locations:[5,6]

$$K(S) = \min\{|s| \text{ s.t. } U(s) = \langle S \rangle\}.$$

The Kolmogorov complexity of a shape $\tilde{S}$ is:

$$K(\tilde{S}) = \min\left\{|s| \text{ s.t. } U(s) = \langle S \rangle \text{ for some } S \in \tilde{S}\right\}.$$

We define the **tile-complexity** of a coordinated shape $S$ and shape $\tilde{S}$ respectively as:

$$K_{sa}(S) = \min\left\{\begin{array}{l} n \text{ s.t. } \exists \text{ a tile system } \mathbf{T} \text{ of } n \text{ tile types that uniquely produces} \\ \text{assembly } A \text{ and } S \text{ is the coordinated shape of } A \end{array}\right\}$$

$$K_{sa}(\tilde{S}) = \min\left\{\begin{array}{l} n \text{ s.t. } \exists \text{ a tile system } \mathbf{T} \text{ of } n \text{ tile types that uniquely produces} \\ \text{assembly } A \text{ and } \tilde{S} \text{ is the shape of } A \end{array}\right\}.$$

**4. Relating Tile-Complexity and Kolmogorov Complexity.** The essential result of this paper is the description of the relationship between the Kolmogorov complexity of any shape and the number of tile types necessary to self-assemble it.

THEOREM 4.1. *There exist constants $a_0, b_0, a_1, b_1$ such that for any shape $\tilde{S}$,*

(4.1) $$a_0 K(\tilde{S}) + b_0 \leq K_{sa}(\tilde{S}) \log K_{sa}(\tilde{S}) \leq a_1 K(\tilde{S}) + b_1.$$

Note that since any tile system of $n$ tile types can be described by $O(n \log n)$ bits, the theorem implies there is a way to construct a tiling system such that asymptotically at least a constant fraction of these bits is used to "describe" the shape rather than any other aspect of the tiling system.

---

[5] Note that $K(S)$ is within an additive constant of $K_U(x)$ where $x$ is some other effective description of $S$, such as a computable characteristic function or a matrix. Since our results are asymptotic, they are independent of the specific representation choice. One might also consider invoking a two dimensional computing machine, but it is not fundamentally different for the same reason.

[6] Notation $\langle \cdot \rangle$ indicates some standard binary encoding of the object(s) in the brackets. In the case of coordinated shapes, it means an explicit binary encoding of the set of locations. Integers, tuples or other data structures are similarly given simple explicit encodings.

*Proof.* [of Theorem 4.1] To see that $a_0 K(\tilde{S}) + b_0 \leq K_{sa}(\tilde{S}) \log K_{sa}(\tilde{S})$, realize that there exists a constant size program $p_{sa}$ that, given a binary description of a tile system, simulates its self-assembly, making arbitrary choices where multiple tile additions are possible. If the self-assembly process terminates, $p_{sa}$ outputs the coordinated shape of the terminal assembly as the binary encoding of the list of locations in it. Any tile system $\mathbf{T}$ of $n$ tile types with any diagonal strength function and any threshold $\tau$ can be represented[7] by a string $d_{\mathbf{T}}$ of $4n\lceil\log 4n\rceil + 16n$ bits: For each tile type, the first of which is assumed to be the seed, specify the bond types on its four sides. There are no more than $4n$ bond types. In addition, for each tile type $t$ specify for which of the 16 subsets $L \subseteq \mathcal{D}$, $\sum_{D \in L} g(bond_D(t)) \geq \tau$. If $\mathbf{T}$ is a tile system uniquely producing an assembly that has shape $\tilde{S}$, then $K(\tilde{S}) \leq |p_{sa}d_{\mathbf{T}}|$. The left inequality in eq. 4.1 follows with the multiplicative constant $a_0 = 1/4 - \varepsilon$ for arbitrary $\varepsilon > 0$.

We prove the right inequality in eq. 4.1 by developing a construction (§5) showing how for any program $s$ s.t. $U(s) = \langle S \rangle$, we can build a tile system $\mathbf{T}$ of $15\frac{|p|}{\log |p|} + b$ tile types, where $b$ is a constant and $p$ is a string consisting of a fixed program $p_{sb}$ and $s$ (i.e. $|p| = |p_{sb}| + |s|$), that uniquely produces an assembly whose shape is $\tilde{S}$. Program $p_{sb}$ and constant $b$ are both independent of $S$. The right inequality in eq. 4.1 follows with the multiplicative constant $a_1 = 15 + \varepsilon$ for arbitrary $\varepsilon > 0$. □

Our result can be used to show that the tile-complexity of shapes is uncomputable:

COROLLARY 4.2. *$K_{sa}$ of shapes is uncomputable. In other words, the following language is undecidable:* $\tilde{L} = \left\{(l,n) \text{ s.t. } l = \langle S \rangle \text{ for some } S \text{ and } K_{sa}(\tilde{S}) \leq n\right\}$. *Language $\tilde{L}$ should be contrasted with $L = \{(l,n) \text{ s.t. } l = \langle S \rangle \text{ and } K_{sa}(S) \leq n\}$ which is decidable (but hard to compute in the sense of NP-completeness [1]).*

*Proof.* [of Corollary 4.2] We essentially parallel the proof that Kolmogorov complexity is uncomputable. If $\tilde{L}$ were decidable, then we can make a program that computes $K_{sa}(\tilde{S})$ and subsequently uses Theorem 4.1 to compute an effective lower bound for $K(\tilde{S})$. Then we can construct a program $p$ that given $n$ outputs some coordinated shape $S$ (as a list of locations) such that $K(\tilde{S}) \geq n$ by enumerating shapes and testing with the lower bound, which we know must eventually exceed $n$. But this results in a contradiction since $p\langle n \rangle$ is a program outputting $S \in \tilde{S}$ and so $K(\tilde{S}) \leq |p| + \lceil \log n \rceil$. But for large enough $n$, $|p| + \lceil \log n \rceil < n$. □

## 5. The Programmable Block Construction.

**5.1. Overview.** The uniquely produced terminal assembly $A$ of our tile system logically will consist of square "blocks" of $c$ by $c$ tiles. There will be one block for each location in $S$. Consider the coordinated shape in Fig. 5.1(a). An example assembly $A$ is graphically represented in Fig. 5.1(b), where each square represents a block containing $c^2$ tiles. Self-assembly initiates in the *seed block*, which contains the seed tile, and proceeds according to the arrows illustrated between blocks. Thus if there is an arrow from one block to another, it indicates that the growth of the second block (a *growth block*) is initiated from the first. A terminated arrow indicates that the block does not initiate the self-assembly of an adjacent block in that direction – in fact, the boundary between such blocks consists of strength-0 interactions (i.e. mismatches). Fig. 5.1(c) describes our nomenclature: an arrow comes into a block on its input side, arrows exit on propagating output sides, and terminated arrows indicate terminating output sides. The seed block has four output sides, which can be either propagating or terminating. Each growth block has one input and three output sides, which are also either propagating or terminating. The overall pattern of bonding of the finished target assembly $A$ is as follows. Tiles on terminal output sides are not bound to the tiles on the adjacent terminal output side (i.e. there is no bonding along the dotted lines in Fig. 5.8(a)), but all other neighboring tiles are bound. We will program the growth such that terminating output sides abut only other terminating output sides or *empty* tiles, and input sides exclusively abut propagating output sides and vice versa.

---

[7] Note that this representation could also be used in the case that negative bond strengths are allowed so long as the strength function is diagonal.

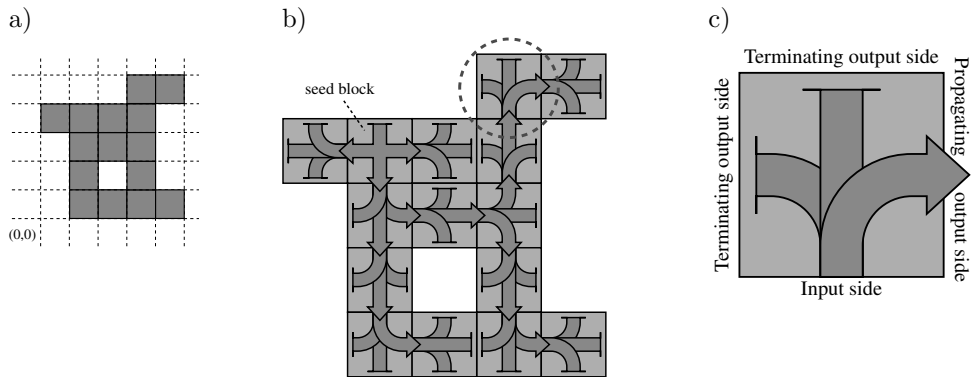

FIG. 5.1. *Forming a shape out of blocks: a) A coordinated shape S. b) An assembly composed of c by c blocks that grow according to transmitted instructions such that the shape of the final assembly is S. Arrows indicate information flow and order of assembly. (Not drawn to scale.) The seed block and the circled growth block are schematically expanded in Fig. 5.2. c) The nomenclature describing the types of block sides.*

The input/output connections of the blocks form a spanning tree rooted at the seed block. During the progress of the self-assembly of the seed block, a computational process determines the input/output relationships of the rest of the blocks in the assembly. This information is propagated from block to block during self-assembly (along the arrows in Fig. 5.1(b)) and describes the shape of the assembly. By following the instructions each growth block receives in its input, the block decides where to start the growth of the next block and what information to pass to it in turn. The scaling factor $c$ is set by the size of the seed block. The computation in the seed block ensures that $c$ is large enough that there is enough space to do the necessary computation within the other blocks.

We present a general construction that represents a Turing-universal way of guiding large scale self-assembly of blocks based on an input program $p$. In the following section, we describe the architecture of seed and growth blocks on which arbitrary programs can be executed. In §5.3 we describe how program $p$ can be encoded using few tile types. In §5.4 we discuss the programming of $p$ that is required to grow the blocks in the form of a specific shape and bound the scaling factor $c$. In §5.5 we demonstrate that the target assembly $A$ is *uniquely* produced.

**5.2. Architecture of the Blocks.**

**5.2.1. Growth Blocks.** There are four types of growth blocks depending upon where the input side is, which will be labeled by $\uparrow$, $\rightarrow$, $\downarrow$ or $\leftarrow$. The internal structure of a $\uparrow$ growth block is schematically illustrated in Fig. 5.2(a). The other three types of growth block are rotated versions of the ↑ block. The specific tile types used for a ↑ growth block are shown in Fig. 5.3, and a simple example is presented in Fig. 5.4. The first part is a Turing machine simulation, which is based on [18, 11]. The machine simulated is a universal Turing machine that takes its input from the propagating output side of the previous block. This TM has an output alphabet $\{0, 1, S\}^3$ and an input alphabet $\{(000), (111)\}$ on a two-way tape (with $\lambda$ used as the blank symbol). The output of the simulation, as 3-tuples, is propagated until the diagonal. The diagonal propagates each member of the 3-tuples crossing it to one of the three output sides, like a prism separating the colors of the spectrum. This allows the single TM simulation to produce three separate strings targeted for the three output sides. The "$S$" symbol in the output of the TM simulation is propagated like the other symbols. However, it acts in a special way when it crosses the boundary tiles at the three output sides of the block, where it starts a new block. The output sides that receive the "$S$" symbol become propagating output sides and the output sides that do not receive it become terminating output sides. In this way, the TM simulation decides which among the three output sides will become propagating output sides, and what information they should contain, by outputting appropriate tuples. Subsequent blocks will use this information as a program, as discussed in §5.4.

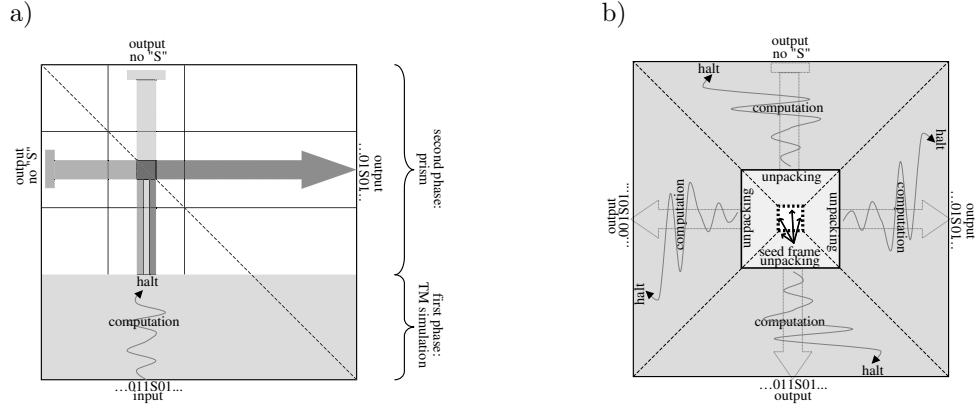

Fig. 5.2. *Internal structure of a growth block (a) and seed block (b).*

**5.2.2. Seed Block.** The internal structure of the seed block is schematically shown in Fig. 5.2(b). It consists of a small square containing all the information pertaining to the shape to be built (the seed frame), a larger square in which this information is unpacked into usable form, and finally four TM simulations whose computations determine the size of the seed block and the information transmitted to the growth blocks. For simplicity we first present a construction without the unpacking process (the *simple* seed block), and then we explain the unpacking process separately and show how it can be used to create the full construction. The tile types used for the simple seed block are presented in Fig. 5.5 and an example is given in Fig. 5.6. While growth blocks contain a single TM simulation that outputs a different string to each of the three output sides, the seed block contains four *identical* TM simulations that output different strings to each of the four output sides. This is possible because the border tile types transmit information selectively: the computation in the seed block is performed using 4-tuples as the alphabet in a manner similar to the growth blocks, but on each side of the seed block only one

a) Borders and basic info propagating tiles:

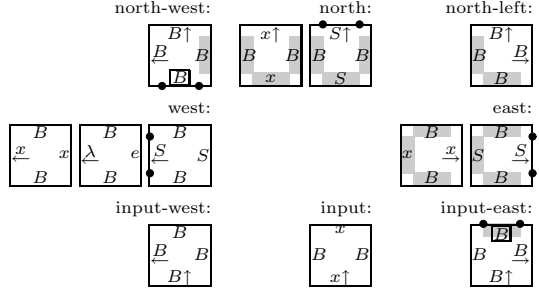

Vertical and horizontal information propagation below the bottom-right/top-left diagonal: $\forall x, y \in \{0, 1, S, \lambda\}^3$:

and above this diagonal: $\forall x, y \in \{0, 1, S, \lambda\}$:

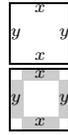

b) Tile types for the diagonal:

TM section diagonal:

Initiation of TM diagonal (to bind to the north-east corner tile) and to delay the upward continuation of the diagonal by one (through the $\delta$ bond):

The prism diagonal, $\forall w, x, y, z \in \{0, 1, S, \lambda\}^3$:

In the row where the Turing machine halts, the $\lambda$ symbol is propagated from the left. This initiates the "prism" diagonal with the following tile:

Termination of the prism diagonal (to bind to the north-west corner tile):

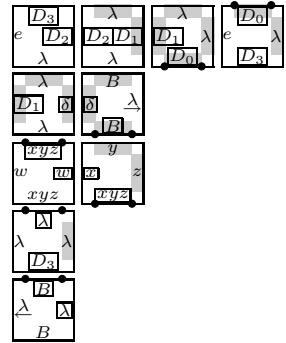

c) TM Simulation tile types:

For every symbol $s$ in $\{0, 1, S, \lambda\}^3$ the following tile types propagate the tape contents: 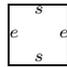

For every symbol $s$ and every state $q$ we add the following "read" tile types: 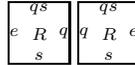

For every symbol $s$ and every state $q$ we add the following "copy" tile type: 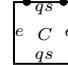

If in state $q$, reading symbol $s$, $U$ writes $s'$, goes to state $q'$, and moves the head right, we add the following "write" tile type: 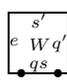

If in state $q$, reading symbol $s$, $U$ writes $s'$, goes to state $q'$, and moves the head left, we add the following "write" tile type: 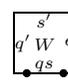

To start $U$ in state $q_0$ we add the following "start" tile type, which places the head at the point at which the "$S$" symbol initiates the block: 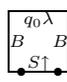

If in state $q$, reading symbol $s$, $U$ halts writing $s'$ then we add the following "halting" tile type: 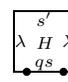

FIG. 5.3. *Growth block $\uparrow$ tile types. All bond types in which a block type symbol is omitted have the block type symbol "$\uparrow$" to prevent inadvertent incorporation of tiles from a different block type. We assume that in bond types above, a single symbol $x \in \{0, 1, S, \lambda\}$ is the same as the tuplet $(xxx)$. The tile types for other growth block types are formed by 90, 180, 270 degree rotations of the tile types of the $\uparrow$ block where the block type symbols $\{\uparrow, \downarrow, \leftarrow, \rightarrow\}$ are replaced by a corresponding 90, 180, 270 degree rotation of the symbol: i.e.* 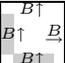 *($\uparrow$ growth block) $\Rightarrow$* 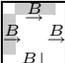 *($\rightarrow$ growth block). Looking at the border tile types, note that external sides of tiles on output sides of blocks have block type symbols compatible with the tiles on an input side of a block. However, tiles on output sides cannot bind to the tiles on an adjacent output side because of mismatching block type symbols.*

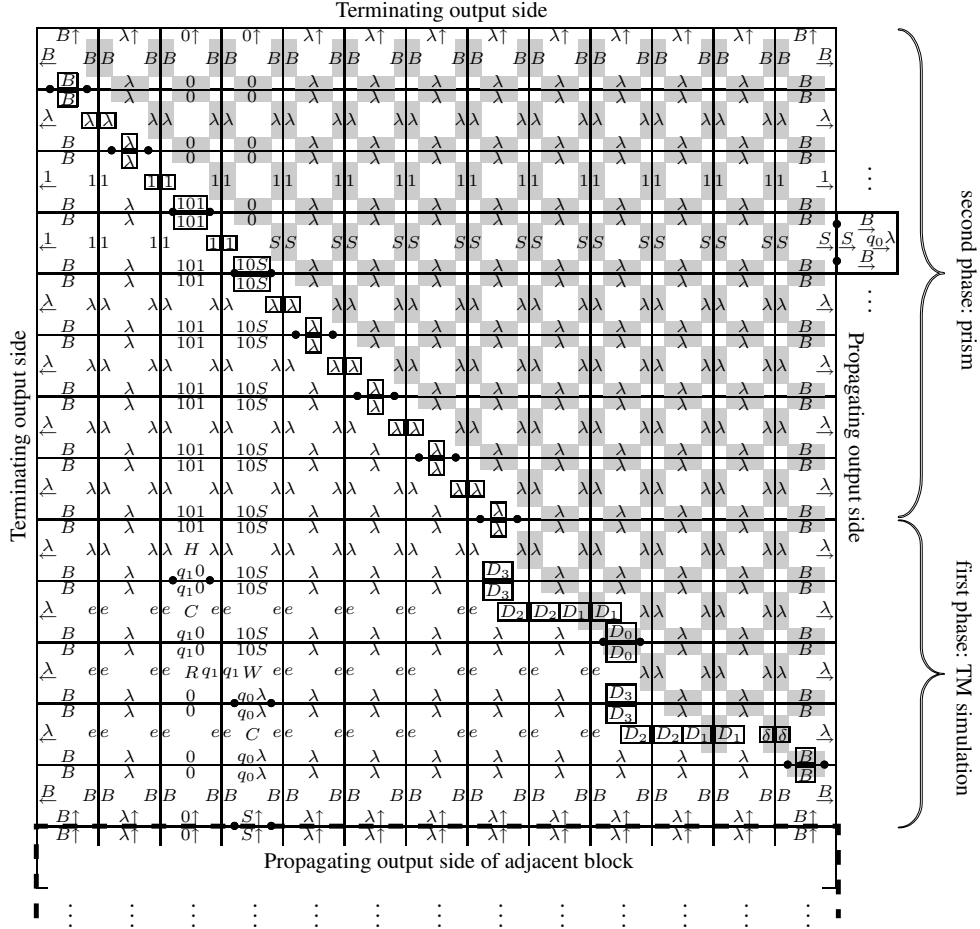

FIG. 5.4. *A trivial example of a ↑ growth block. Here, the TM makes one state transition and halts. All bond types in which a block type symbol is omitted have the block type symbol "↑". We assume that in bond types above, a single symbol $x \in \{0, 1, S, \lambda\}$ is the same as the tuplet $(xxx)$. The natural assembly sequence to consider is adding tiles row by row from the south side (in which a new row is started by the strength 2 bond).*

of the elements of the 4-tuple traverses the border. As with growth blocks, if the transmitted symbol is "S", the outside edge initiates the assembly of the adjoining block. The point of having four identical TM simulations is to ensures that the seed block is square: while a growth block uses the length of its input side to set the length of its output sides (via the diagonal), the seed block does not have any input sides. (Remember that it is the seed block that sets the size of all the blocks.)

The initiation of the Turing machine simulations in the seed block is done by tile types encoding the program $p$ that guides the block construction. The natural approach to provide this input is using 4 rows (one for each TM) of unique tiles encoding one bit per tile, as illustrated in Figs. 5.5 and 5.6. However, this method does not result in an asymptotically optimal encoding.

**5.3. The Unpacking Process.** To encode bits much more effectively we follow Adleman et al [3] and encode on the order of $\log n / \log \log n$ bits per tile where $n$ is the length of the input. This representation is then unpacked into a one-bit-per-tile representation used by the TM simulation. Adleman et al's method requires $O(n/\log n)$ tiles to encode $n$ bits, leading to the asymptotically optimal result of Theorem 4.1.

Our way of encoding information is based on Adleman et al [3], but modified to work in a $\tau = 2$ tile system (with strength function ranging over $\{0, 1, 2\}$) and to fit our construction in its geometry. We express a length $n$ binary string using a concatenation of $\lceil n/k \rceil$ binary substrings of length $k$, padding with 0's if necessary.[8] We choose $k$ such that it is the least integer satisfying $\frac{n}{\log n} \leq 2^k$. Clearly, $2^k < \frac{2n}{\log n}$. See Fig. 5.7 for the tile types used in the unpacking for the north TM simulation and for a simple unpacking example (which for the sake of illustration uses $k = 4$).

Let us consider the number of tile types used to encode and unpack the $n$ bit input string for a single TM simulation (i.e. north). There are $2\lceil n/k \rceil \leq 2\lceil \frac{n}{\log \frac{n}{\log n}} \rceil = 2\lceil \frac{n}{\log n - \log \log n} \rceil$ unique tile types in each seed row. This implies that there exists a constant $h$ such that $2\lceil n/k \rceil \leq \frac{3n}{\log n} + h$ for all $n$. We need at most $2^k + 2^{k-1} + \cdots + 4 < 2^{k+1}$ "extract bit" tile types and $2^{k-1} + 2^{k-2} + \cdots + 4 < 2^k$ "copy remainder" tile types. To initiate the unpacking of new substrings we need $2^k$ tile types. To keep on copying substrings that are not yet unpacked we need $2(2^k)$ tile types. The quantity of the other tile types is independent of $n, k$. Thus, in total, to unpack the $n$ bit input string for a single TM simulation we need no more than $\frac{3n}{\log n} + h + 2^{k+1} + 2^k + 2^k + 2(2^k) \leq 15\frac{n}{\log n} + O(1)$ tile types. Since there are 4 TM simulations in the seed block, we need $60\frac{n}{\log n} + O(1)$ tile types to encode and unpack the $n$ bit input string.

If the seed block requires only one propagating output side, then a reduced construction using fewer tile types can be used: only one side of the seed frame is specified, and only one direction of unpacking tiles are used. A constant number of additional tile types are used to fill out the remaining three sides of the square. These additional tile types must perform two functions. First, they must properly extend the diagonal on either side of the unpacking and TM simulation regions. In the absense of the other three unpacking and TM simulation processes, this requires adding strength-2 bonds that allow the diagonal to grow to the next layer. Second, the rest of the square must be filled in to the correct size. This can be accomplished by adding tiles that extend one diagonal to the other side of the seed frame (using the same logic as a construction in [11].) Altogether, a seed block with only one propagating output side

---

[8]We can assume that our universal TM U treats trailing 0's just as $\lambda$'s

a) Borders and half-diagonals:

The borders:
$\forall w, x, y, z \in \{0, 1, \lambda\}$:

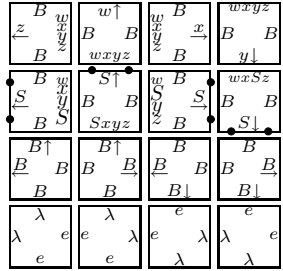

Corner tile types:

The four half-diagonals to separate the TM simulations and augment the TM tape with blanks:

b) Seed frame for program $p$.

TM seed frame: for every symbol $p_i$:

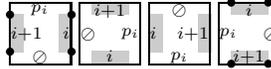

If $p_i$ is "$U$" then the corresponding bond type is strength 2, starting the TM simulation with the head positioned at that point reading $\lambda$.

Corners of the seed frame: let $i_m = |p|$:

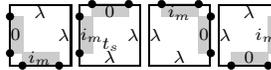

We make the north-west corner the seed tile of our tile system.

To fill in the middle:

c) TM Simulation tile types (north only):

For every symbol $s$ in $\{0, 1, S, \lambda\}^4$ the following tile types propagate the tape contents: 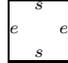

For every symbol $s$ and every state $q$ we add the following "read" tile types: 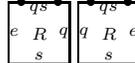

If in state $q$, reading symbol $s$, $U$ writes $s'$, goes to state $q'$, and moves the head left, we add the following "write" tile type: 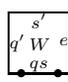

If in state $q$, reading symbol $s$, $U$ writes $s'$, goes to state $q'$, and moves the head right, we add the following "write" tile type: 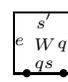

To start $U$ in state $q_0$ we add the following "start" tile type, which places the head at the point at which the "$S$" symbol initiates the block: 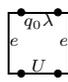

If in state $q$, reading symbol $s$, $U$ halts writing $s' = (wxyz)$ then we add the following "halting" tile type, which also starts the border: 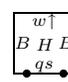

FIG. 5.5. *Seed block tile types without unpacking. All bond types in which a block type symbol is omitted have the block type symbol "☆" to prevent inadvertent incorporation of tiles from a different block type. We assume that in bond types above, a single symbol $x \in \{0, 1, S, \lambda\}$ is the same as the tuplet $(xxxx)$. Note that as with output sides of growth blocks, the external sides of seed block border tiles have block type symbols compatible with the tiles on an input side of a growth block. The three other TM simulations consist of tile types that are rotated versions of the north TM simulation shown. The halting tile types propagate one of the members of the tuple on which the TM halts, analogous to the border tile types. The bond types of TM tile types have a symbol from $\mathcal{D}$ which indicates which simulation they belong to (omitted above).*

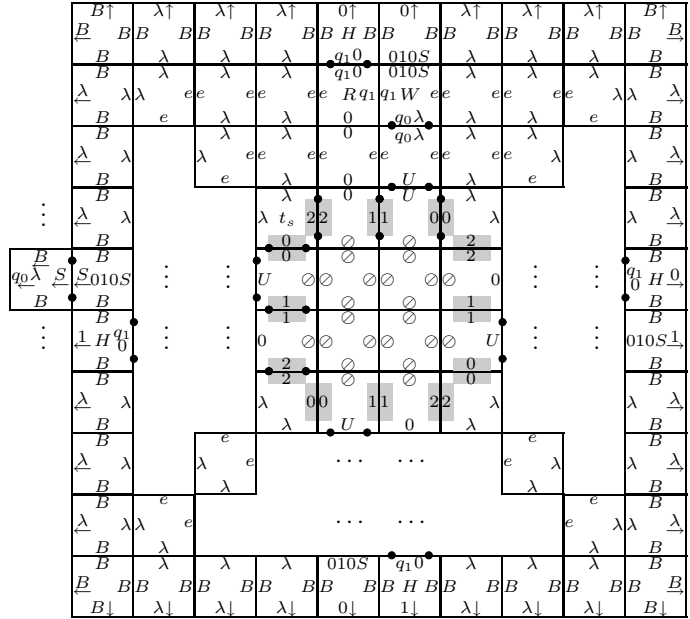

FIG. 5.6. *A simple seed block without unpacking showing the north TM simulation and the selective transmission of information through the borders. As shown, only the west side is a propagating output side; the other three sides are terminating output sides. All bond types in which a block type symbol is omitted have the block type symbol "☆". We assume that in bond types above, a single symbol $x \in \{0, 1, S, \lambda\}$ is the same as the tuplet $(xxxx)$. The natural assembly sequence to consider is growing the seed frame first and then adding tiles row by row from the center (where a new row is started by the strength 2 bond).*

a) Unpacking tile types for the north side of the seed frame:

We use $n/k$ coding tiles in the input row, each encoding a binary substring ($w_i$) of length $k$. These tiles are interspersed with buffer tiles holding the symbol "$*$". $\forall 0 \geq i \geq k/n - 1$:

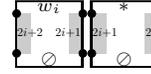

The last tile of the seed row has symbol "$U$" which indicates the end of the input string.

To initiate the unpacking of new substrings: $\forall x \in \{0,1\}^{k-1}$, $b \in \{0,1\}$:

The following "extract bit" tile types perform the actual unpacking: $\forall j \in \{1, \ldots, k-1\}, \forall x \in \{0,1\}^j, b \in \{0,1\}$:

The following "copy remainder" tile types pass the remaining bits to the next extraction: $\forall j \in \{2, \ldots, k-1\}, \forall x \in \{0,1\}^j$:

To copy a single bit in the last step of the unpacking of a substring and after unpacking every bit: $b \in \{0,1\}$:

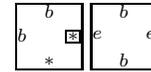

These tile types keep on copying substrings that are not yet being unpacked: $\forall x \in \{0,1\}^k$:

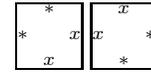

Finally, the following tile types propagate the symbol "$U$", which indicates the end of the input string, and initiate the TM simulation once the unpacking process finishes:

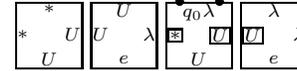

b) North unpacking example:

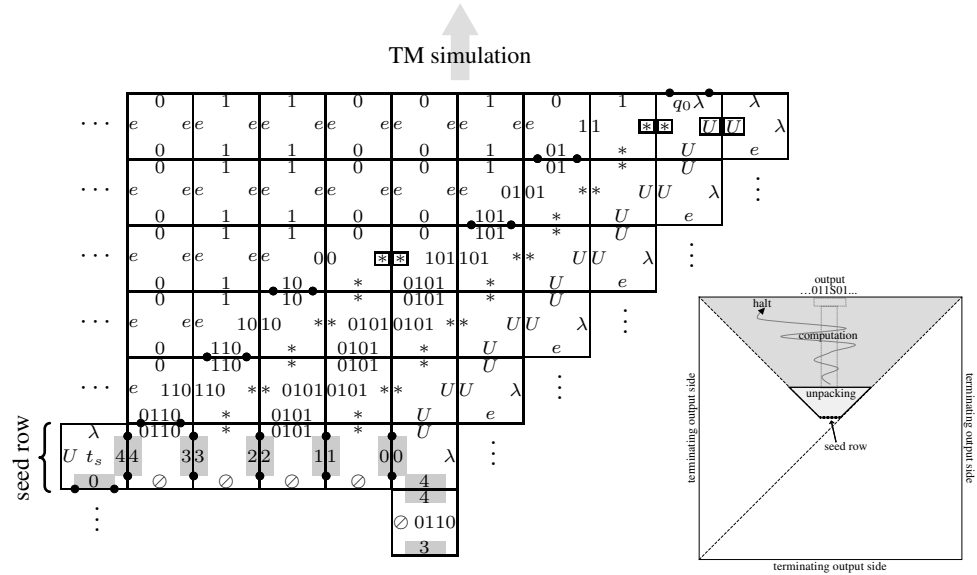

FIG. 5.7. *The unpacking for the north side of the seed frame. (a) The tile types used. (b) An example showing the unpacking of the string* 01100101 *if* $k = 4$ *for a seed block with up to four propagating output sides. Note that the unpacking process can be inserted immediately prior to the TM simulation without modifying other tile types. (inset) Internal structure of a seed block with only one propagating output side.*

requires only $15\frac{n}{\log n} + O(1)$ tile types. We will see in the next section that this is sufficient for growing any shape.

**5.4. Programming Blocks and the Value of the Scaling Factor $c$.** In order for our tile system to produce some assembly whose shape is $\tilde{S}$, instructions encoded in $p$ must guide the construction of the blocks by deciding on which side of which block a new block begins to grow and what is encoded on the edge of each block. For our purposes, we take $p = p_{sb}\langle s \rangle$ (i.e. $p_{sb}$ takes $s$ as input), where $s$ is a program that outputs the list of locations in the shape $S$. $p_{sb}$ runs $s$ to obtain this list and plans out a spanning tree $t$ over these locations (it can just do a depth-first search) starting from some arbitrarily chosen location that will correspond to the seed block.[9] The information passed along the arrows in Fig. 5.1(b) is $p_{gb}\langle t, (i,j) \rangle$ which is the concatenation of a program $p_{gb}$ to be executed within each growth block, and an encoding of the tree $t$ and the location $(i,j)$ of the block into which the arrow is heading. When executed, $p_{gb}\langle t, (i,j) \rangle$ evaluates to a 3-tuple encoding of $p_{gb}\langle t, D(i,j) \rangle$ together with symbol "$S$" for each propagating output side $D$. Thus, each growth block passes $p_{gb}\langle t, D(i,j) \rangle$ to its $D^{th}$ propagating output side as directed by $t$. Note that program $p_{sb}$ in the seed tile must also run long enough to ensure that $c$ is large enough that the computation in the growth blocks has enough space to finish without running into the sides of the block or into the diagonal. Nevertheless, the scaling factor $c$ is dominated by the building of $t$ in the seed block, as the computation in the growth blocks takes only $poly(|S|)$.[10] Since the building of $t$ is dominated by the running time of $s$, we have $c = poly(time(s))$.

**5.5. Uniqueness of the Terminal Assembly.** By Theorem 2.3 it is enough to demonstrate a locally deterministic assembly sequence ending in our target terminal assembly to be assured that this terminal assembly is uniquely produced. Consider the assembly sequence $\vec{A}$ in which the assembly is constructed *block by block* such that every block is finished before the next one started and each block is constructed by the natural assembly sequence described in the captions to Figs. 5.4 and 5.6. It is enough to confirm that in this natural assembly sequence every tile addition satisfies the definition of local determinism (Def. 2.2). It is easy to confirm that every tile not adjacent to a terminal output side of a block indeed satisfies these conditions. Other than on a terminal output side of a block (and on *null* tiles) there are no *termsides*: every side is either an *inputside* or a *propside*. In our construction, each new tile binds through either a single strength 2 bond or two strength 1 bonds (thus condition 1 is satisfied since $\tau = 2$) such that no other tile type can bind through these *inputsides* (condition 2 is satisfied if the tile has no *termsides*). Note that inadvertent binding of a tile type from a different block type is prevented by the block type symbols.

Now let's consider *termsides* around the terminal output sides of blocks (Fig. 5.8(a)). Here block type symbols come to the rescue again and prevent inadvertent binding. Let $t \in A$ be a tile with a *termside* ($t$ can be *null*). We claim that $\forall t'$ s.t. $type(t') \in T$ and $pos(t') = pos(t)$, if $\Gamma^A_{termsides^{\vec{A}}(t)}(t') > 0$ then $\Gamma^A_{\mathcal{D}-propsides^{\vec{A}}(t)}(t') < \tau = 2$. In other words, if $t'$ binds on a *termside* of $t$, then it cannot bind strongly enough to violate local determinism, implying we can ignore *termsides*. Figure 5.8(a) shows in dotted lines the *termsides* that could potentially be involved in bonding. These *termsides* cannot have a strength 2 bond because symbol "$S$" is not propagated to terminal output sides of blocks. Thus $t'$ binding only on a single *termside* of $t$ is not enough. Can $t'$ bind on two *termsides* of $t$? To do so, it must be in a corner between two blocks, binding two terminal output sides of different blocks. But to bind in this way would require $t'$ to bond the block type symbol pattern[11] shown in Fig. 5.8(b) (or its rotation), which none of the tile types in our tile system can do. Can $t'$ bind on one *termside* and one *inputside* of $t$? Say the *termside* of $t$ that $t'$ binds on is the west side (Fig. 5.8(c)). The tile to the west of $t$ must be on the east terminal output side of a block, and thus it has symbol "$\rightarrow$" on its east side. So $t'$ must have "$\rightarrow$" on the west, and depending on the type of block $t$ is in, one of the other block type symbols as shown in Fig. 5.8(c). But again none of the tile types in our tile system has the necessary block type symbol pattern.

---

[9] We can opt to always choose a leaf, in which case the seed block requires only one propagating output side. In this case the multiplicative factor $a_1$ is $15 + \varepsilon$, although the tile set used will depend upon the direction of growth from the leaf.

[10] Note that less than $n$ rows are necessary to unpack a string of length $n$ (§5.3). Since we can presume that $p_{sb}$ reads its entire input and the universal TM needs to read the entire input program to execute it, the number of rows required for the unpacking process can be ignored with respect to the asymptotics of the scaling factor $c$.

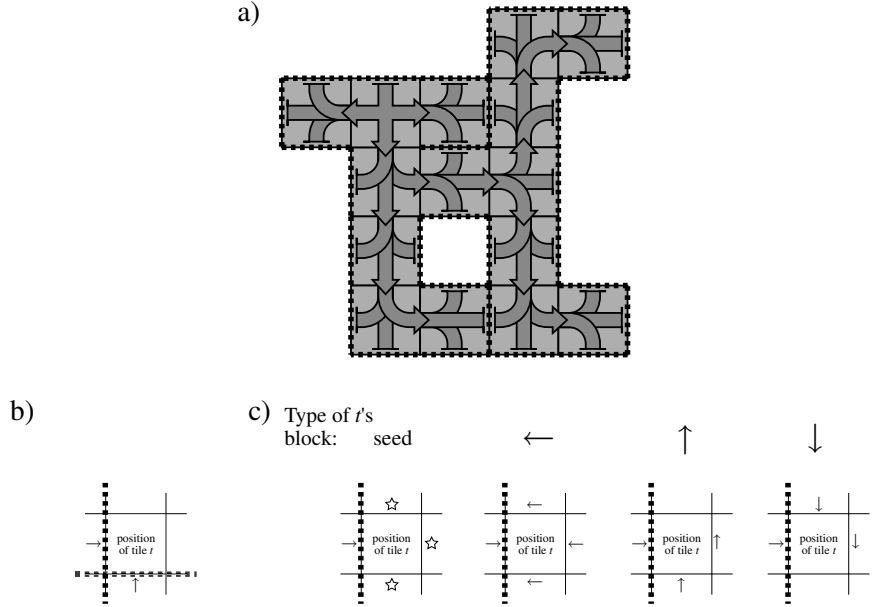

FIG. 5.8. *(a) The target terminal assembly with the dotted lines indicating the edges that have termsides with non-null bonds. (b) The block type symbols of adjacent tiles on two termsides of $t$ (west and south in this case). (c) The block type symbols of adjacent tiles on a termside (west side in this case) and an inputside of $t$. If $t$ is in the seed block or $\leftarrow$ growth block, then the north, east, south sides may be the inputsides. If $t$ is in a $\uparrow$ block then the east and south sides may be the inputsides. If $t$ is in a $\downarrow$ block then the north and east sides may be the inputsides.*

---

[11]The block type symbol pattern of a tile type consists of the block type symbols among its four bond types. For instance, the tile type $\begin{smallmatrix} \lambda\uparrow \\ \lambda\uparrow \phantom{x} \lambda\uparrow \\ D_3\uparrow \end{smallmatrix}$ has block type symbol pattern $\begin{smallmatrix} \uparrow \\ \uparrow \phantom{x} \uparrow \\ \uparrow \end{smallmatrix}$. If two bond types do not have matching block type symbols then obviously they cannot bind.

**6. Discussion, Extensions and Open Questions.** In this work we have established both upper and lower bounds relating the descriptional complexity of a shape to the number of tile types needed to self-assemble the shape within the standard Tile Assembly Model. The relationship is dependent upon a particular definition of shape that ignores its size. Disregarding scale in self-assembly appears to play a similar role as disregarding time in theories of computability and decidability. Those theories earned their universal standing by being shown to be identical for all "reasonable" models of computation. To what extent do our results depend on the particular model of self-assembly? Can one define a complexity theory for families of shapes in which the absolute scale is the critical resource being measured? In this section we discuss the generality and limitations of our result.

**6.1. Optimizing the Main Result (§4).** Since the Kolmogorov complexity of a string depends on the universal Turing machine chosen, the complexity community adopted a notion of additive equivalence, where additive constants are ignored. However, Theorem 4.1 includes multiplicative constants as well, which are not customarily discounted. It might be possible to use a more clever method of unpacking (§5.2) and a seed block construction that reduces the multiplicative constant $a_1$ of Theorem 4.1. Correspondingly, there might be a more efficient way to encode any tile system than described in the proof of the theorem, and thereby increase $a_0$.

Recall that $s$ is the program for $U$ producing the target coordinated shape $S$ as a list of locations. For cases where our results are of interest, the scaling factor $c = O(time(s))$ is extremely large since $|S|$ is presumably enormous and $s$ must output every location in $S$. The program $s'$ that given $(i, j)$ outputs 0/1 indicating whether $S$ contains that location may run much faster than $s$ for large shapes. Can our construction be adapted to use $s'$ in each block rather than $s$ in the seed block to obtain smaller scale? The problem with doing this directly is that the scale of the blocks, which sets the maximum allowed running time of computation in each block, must be set in the seed block. As a result, there must be some computable time bound on $s'$ that is given to the seed block.

**6.2. Strength Functions.** In most previous works on self-assembly, as in this work, strength functions are restricted with the following properties: (1) the effect that one tile has on another is equal to the effect that the other has on the first, i.e., $g$ is *symmetric*: $g(\sigma, \sigma') = g(\sigma', \sigma)$; (2) the lack of an interaction is normalized to zero, i.e., $g(\sigma, null) = 0$; (3) there are no "adverse" interactions counteracting other interactions, i.e., $g$ is *non-negative*; (4) only sides with matching bond types interact, i.e., $g$ is *diagonal*: $g(\sigma, \sigma') = 0$ if $\sigma \neq \sigma'$.

Properties 1 and 2 seem natural enough. Our results are independent of property 3 because the encoding used for the lower bound of Theorem 4.1 is valid for strength functions taking on negative values. Property 4, which reflects the roots of the Tile Assembly Model in the Wang tiling model, is essential for the quantitative relationship expressed in Theorem 4.1: recent work by Aggarwal et al [4] shows that permitting non-diagonal strength functions allows information to be encoded more compactly. Indeed, if property 4 is relaxed then replacing our unpacking process with the method of encoding used in that work and using Aggarwal et al's lower bound leads to the following form of Theorem 4.1: Assuming the maximum threshold $\tau$ is bounded by a constant, there exist constants $a_0, b_0, a_1, b_1$ such that for any shape $\tilde{S}$,

$$a_0 K(\tilde{S}) + b_0 \leq \left(K_{sa}^{nd}(\tilde{S})\right)^2 \leq a_1 K(\tilde{S}) + b_1$$

where $K_{sa}^{nd}$ is the tile-complexity when non-diagonal strength functions are allowed. It is an open question whether the constant bound on $\tau$ can be relaxed.

**6.3. Wang Tiling vs Self-Assembly of Shapes.** Suppose one is solely concerned with the existence of a configuration in which all sides match, and not with the process of assembly. This is the view of classical tiling theory [7]. Since finite tile sets can enforce uncomputable tilings of the plane [8, 16], one might expect greater computational power when the existence, rather than production, of a tiling is used to specify shapes. In this section we develop the notion of shapes in the Wang tile model [19] and show that results almost identical to the Tile Assembly Model hold. One conclusion of this analysis is that making a shape "practically constructible" (ie in the sense of the Tile Assembly Model) does not necessitate an increase in tile-complexity.

We translate the classic notion of the origin-restricted Wang tiling problem[12] as follows. A **(origin-restricted) Wang tiling system** is a pair $(T, t_s)$ where $T$ is a set of tile types and $t_s$ is a **seed tile** with $type(t_s) \in T$. A configuration $A$ is a valid tiling if all sides match and it contains the seed tile. Formally, $A$ is a **valid tiling** if $\forall (i,j) \in \mathbb{Z}^2, D \in \mathcal{D}$, (1) $type(A(i,j)) \in T$, (2) $t_s \in A$, (3) $bond_D(A(i,j)) = bond_{D^{-1}}(A(D(i,j)))$.

Since valid tilings are infinite objects, how can they define finite coordinated shapes? For tile sets containing the *empty* tile type, we can define shapes analogously to the Tile Assembly Model. However, we cannot simply define the coordinated shape of a valid tiling to be the set of locations of non-empty tiles. For one thing, the set of non-empty tiles can be disconnected, unlike in self-assembly where any produced assembly is a single connected component. So we take the coordinated shape $S_A$ of a valid tiling $A$ to be the smallest region of non-*empty* tiles containing $t_s$ that can be extended to infinity by *empty* tiles. Formally, $S_A$ is the coordinated shape of the smallest subset of $A$ that is a valid tiling containing $t_s$. If $S_A$ is finite, then it is the **coordinated shape of valid tiling** $A$.[13] Shape $\tilde{S}$ is the **shape of a valid tiling** $A$ if $S_A \in \tilde{S}$.

Produced assemblies of a tile system $(T, t_s, g, \tau)$ are not necessarily valid tilings of Wang tiling system $(T, t_s)$ because the Tile Assembly Model allows mismatching sides. Further, valid tilings of $(T, t_s)$ are not necessarily produced assemblies of $(T, t_s, g, \tau)$. Even if one considers only valid tilings that are connected components, there might not be any sequence of legal tile additions that assembles these configurations. Nonetheless, if a tile system uniquely produces a valid tiling $A$, then all valid tilings of the corresponding Wang tile system agree with $A$ and have the same coordinated shape as $A$:

LEMMA 6.1. *If empty $\in T$ and the tile system* $\mathbf{T} = (T, t_s, g, \tau)$ *uniquely produces assembly $A$ such that $A$ is a valid tiling of the Wang tiling system $(T, t_s)$ then for all valid tilings $A'$: (1) $\forall (i,j) \in \mathbb{Z}^2$, $type(A(i,j)) \neq empty \Rightarrow A'(i,j) = A(i,j)$, (2) $S_{A'} = S_A$.*

*Proof.* Consider an assembly sequence $\vec{A}$ of $\mathbf{T}$ ending in assembly $A$ and let $A'$ be a valid tiling of $(T, t_s)$. Suppose $t_n$ is the first tile added in this sequence such that $t' = A'(pos(t_n)) \neq t_n$. Since $A'$ is a valid tiling, $t'$ must match on all sides, including $inputsides^{\vec{A}}(t_n)$. But this implies that two different tiles can be added in the same location in $\vec{A}$ which means that $A$ is not uniquely produced. This implies part (1) of the lemma. Now, to be a valid tiling, all exposed sides of assembly $A$ must be null. Thus if $A'$ and $A$ agree on all places where $A$ is non-*empty*, then $S_{A'} = S_A$ and part (2) of the lemma follows. □

---

[12]The *unrestricted* Wang tile model does not have a seed tile [19, 5, 18].

[13]$S_A$ can be finite only if $empty \in T$ because otherwise no configuration containing an *empty* tile can be a valid tiling.

Define the **tile-complexity** $K_{wt}$ of a shape $\tilde{S}$ in the origin-restricted Wang tiling model as the minimal number of tile types in a Wang tiling system with the property that a valid tiling exists and there is a coordinated shape $S \in \tilde{S}$ such that for every valid tiling $A$, $S_A = S$.

THEOREM 6.1. *There exist constants $a_0, b_0, a_1, b_1$ such that for any shape $\tilde{S}$,*

$$a_0 K(\tilde{S}) + b_0 \leq K_{wt}(\tilde{S}) \log K_{wt}(\tilde{S}) \leq a_1 K(\tilde{S}) + b_1.$$

*Proof.* [Sketch] The left inequality follows in a manner similar to the proof of Theorem 4.1. Suppose every valid tiling of our Wang tile system has coordinated shape $S$. Any Wang tiling system of $n$ tile types can be represented using $O(n \log n)$ bits. Making use of this information as input, we can use a constant-size program to find, through exhaustive search, the smallest region containing $t_s$ surrounded by *null* bond types in some valid tiling. Thus, $O(n \log n)$ bits are enough to compute an instance of $\tilde{S}$. To prove the right inequality, our original block construction almost works, except that there are mismatches between a terminal output side of a block and the abutting terminal output side of the adjacent block or the surrounding *empty* tiles (i.e. along the dotted lines in Fig. 5.8(a)). Consequently, the original construction does not yield a valid tiling. Nonetheless, a minor variant of our construction overcomes this problem. Instead of relying on mismatching bond type symbols to prevent inadvertent binding to terminal output sides of blocks, we can add an explicit capping layer that covers the terminal output sides with *null* bond types but propagates information through propagating output sides. This way, the terminal output sides of blocks are covered by *null* bond types and match the terminal output sides of the adjacent block and *empty* tiles. These modifications can be made preserving local determinism, which, by Lemma 6.1, establishes that the coordinated shape of any valid tiling is an instance of $\tilde{S}$. □

There may still be differences in the computational power between Wang tilings and self-assembly processes. For example, consider the smallest Wang tiling system and the smallest self-assembly tile system that produce instances of $\tilde{S}$. The instance produced by the Wang tiling system might be much smaller than the instance produced by self-assembly.

Keep in mind that the definition we use for saying when a Wang tiling system produces a shape was chosen as a natural parallel to the definition used for self-assembly, but alternative definitions may highlight other interesting phenomena specific to Wang tilings. For example, one might partition tiles into two subsets, "solution" and "substance" tiles, and declare shapes to be connected components of substance tiles within valid tilings. In such tilings – reminiscent of "vicinal water" in chemistry – the solution potentially can have a significant (even computational) influence that restricts possible shapes of the substance, and hence the size of produced shapes needn't be so large as to contain the full computation required to specify the shape.

**6.4. Sets of Shapes.** Any coordinated shape $S$ can be trivially produced by a self-assembly tile system or by a Wang tiling of $|S|$ tile types. Interesting behavior occurs only when the number of tile types is somehow restricted and the system is forced to perform some non-trivial computation to produce a shape. Previously in this paper, we restricted the number of tile types in the sense that we ask what is the minimal number of tile types that can produce a given shape. We saw that ignoring scale in this setting allows for an elegant theory. In the following two sections the restriction on the number of tile types is provided by the infinity of shapes they must be able to produce. Here we will see as well that ignoring scale allows for an elegant theory.

Adleman [2] asks "What are the 'assemblable [sic] shapes?' - (analogous to what are the 'computable functions')?" While this is still an open question for coordinated shapes, our definition of a shape ignoring scale and translation leads to an elegant answer.

A set of binary strings $\tilde{L}$ is a language of shapes if it consists of (standard binary) encodings of lists of locations that are coordinated shapes in some set of shapes: $\tilde{L} = \left\{ \langle S \rangle \text{ s.t. } S \in \tilde{S} \text{ and } \tilde{S} \in R \right\}$ for some set of shapes $R$. Note that every instance of every shape in $R$ is in this language. The language of shapes $\tilde{L}$ is recursively enumerable if there exists a Turing machine that halts upon receiving $\langle S \rangle \in \tilde{L}$, and does not halt otherwise. We say a tile system $\mathbf{T}$ produces the language of shapes $\tilde{L}$ if $\tilde{L} = \left\{ \langle S \rangle \text{ s.t. } S \in \tilde{S}_A \text{ for some } A \in Term(\mathbf{T}) \right\}$. We may want $\tilde{L}$ to be *uniquely produced* in the sense that the $A \in Term(\mathbf{T})$ is unique for each shape. Further, to prevent infinite spurious growth we may also require $\mathbf{T}$ to satisfy the *non-cancerous* property: $\forall B \in Prod(\mathbf{T}), \exists A \in Term(\mathbf{T})$ s.t. $B \to_\mathbf{T}^* A$. The following lemma is valid whether or not these restrictions are made.

LEMMA 6.2. *A language of shapes $\tilde{L}$ is recursively enumerable if and only if it is (uniquely) produced by a (non-cancerous) tile system.*

*Proof.* [Sketch] First of all, for any tile system $\mathbf{T}$ we can make a TM that given a coordinated shape $S$ as a list of locations, starts simulating all possible assembly sequences of $\mathbf{T}$ and halts iff it finds a terminal assembly that has shape $\tilde{S}$. Therefore, if $\tilde{L}$ is produced by a tile system, $\tilde{L}$ is recursively enumerable. In the other direction, if $\tilde{L}$ is recursively enumerable then there is a program $p$ that given $n$ outputs the $n^{th}$ shape from $\tilde{L}$ (in some order) without repetitions. Our programmable block construction can be modified to execute a non-deterministic universal TM in the seed block by having multiple possible state transitions. We make a program that non-deterministically guesses $n$, feeds it to $p$, and proceeds to build the returned shape. Note that since every computation path terminates, this tile system is non-cancerous, and since $p$ enumerates without repetitions, the language of shapes is uniquely produced. □

Note that the above lemma does not hold for languages of coordinated shapes, defined analogously. Many simple recursively enumerable languages of coordinated shapes cannot be produced by any tile system. For example, consider the language of equilateral width-1 crosses centered at $(0, 0)$. No tile system produces this language. Scale equivalence is crucial because it allows arbitrary amounts of information to be passed between different parts of a shape; otherwise, the amount of information is limited by the width of a shape.

The same lemma can be attained for the Wang tiling model in an analogous manner using the construction from §6.3. Let us say a Wang tiling system $(T, t_s)$ produces the language of shapes $\tilde{L}$ if $\tilde{L} = \{\langle S \rangle \text{ s.t. } S \in \tilde{S}_A \text{ for some valid tiling } A \text{ of } (T, t_s)\}$. Analogously to tile systems, we may require the *unique production* property that there is exactly one such $A$ for each shape. Likewise, corresponding to the non-cancerous property of tile systems, we may also require the tiling system to have the *non-cancerous* property that every valid tiling has a coordinated shape (i.e. is finite).

Again, the following lemma is true whether or not these restrictions are made.

LEMMA 6.3. *A language of shapes $\tilde{L}$ is recursively enumerable if and only if it is (uniquely) produced by a (non-cancerous) Wang tiling system.*

**6.5. Scale Complexity of Shape Functions.** In this section we expand upon the notion of a shape being the output of a universal computation process as mentioned in the Introduction. Here we will think of tile systems as computing a function from binary strings to shapes and show that the time complexity of this function in terms of Turing machines is closely related to the total number of tiles used to assemble the output shape (not tile *types*). The equivalent connection can be made between non-deterministic Turing machines and the size of valid tilings in the Wang tiling model.

Let $f$ be a function from binary strings to shapes. We say that a Turing machine $M$ computes this function if for all $x$, $f(x) = \tilde{S} \Leftrightarrow \exists S \in \tilde{S}$ s.t. $M(x) = \langle S \rangle$. The standard notion of time-complexity applies: $f \in TIME_{TM}(t(n))$ if there is a TM computing it running in time bounded by $t(n)$ where $n$ is the size of the input. In §5.2.2 we saw how binary input can be provided to a tile system via a seed frame wherein all four sides of a square present the bitstring. Let us apply this convention here.[14] Extending the notion of the seed in self-assembly to the entire seed frame and using this as the input for a computation[17], we say a tile system computes $f$ if: [starting with the seed frame encoding $x$ the tile system uniquely produces an assembly of shape $\tilde{S}$] iff $f(x) = \tilde{S}$. We say that $f \in TILES_{SA}(t(n))$ if there is a tile system computing it and the size of coordinated shapes produced (in terms of the number of non-empty locations) for inputs of size $n$ is upper bounded by $t(n)$. Similar definitions can be made for non-deterministic Turing machines and Wang tiling systems. We say that a NDTM $N$ computes $f$ if: [every computation path of $N$ on input $x$ ending in an accept state (as opposed to a reject state) outputs $\langle S \rangle$] iff $f(x) = \tilde{S}$. For non-deterministic Turing machines, $f \in TIME_{NDTM}(t(n))$ if there is a NDTM computing $f$ such that every computation path halts after $t(n)$ steps. Extending the notion of the seed for Wang tilings to the entire seed frame as well, we say a Wang tiling system computes $f$ if: all valid tilings containing the seed frame have a coordinated shape and this coordinated shape is the same for all such valid tilings, and it is an instance of the shape $f(x)$. We say that $f \in TILES_{WT}(t(n))$ if there is a tiling system computing it and the size of coordinated shapes produced for inputs of size $n$ is upper bounded by $t(n)$.

THEOREM 6.4.
(a) If $f \in TILES_{SA}(t(n))$ then $f \in TIME_{TM}(O(t(n)^4))$
(b) If $f \in TIME_{TM}(t(n))$ then $f \in TILES_{SA}(O(t(n)^3))$
(c) If $f \in TILES_{WT}(t(n))$ then $f \in TIME_{NDTM}(O(t(n)^4))$
(d) If $f \in TIME_{NDTM}(t(n))$ then $f \in TILES_{WT}(O(t(n)^3))$

*Proof.* [Sketch] (a) Let **T** be a tile system computing $f$ such that the total number of tiles used on an input of size $n$ is $t(n)$. A Turing machine with a 2D tape can simulate the self-assembly process of **T** with an input of size $n$ in $O(t(n)^2)$ time: for each of the $t(n)$ tile additions, it needs to search $O(t(n))$ locations for the next addition. This 2D Turing machine can be simulated by a regular Turing machine with a quadratic slowdown.[15]

---

[14] Any other similar method would do. For the purposes of this section, it does not matter whether we use the one bit per tile encoding or the encoding requiring unpacking (§5.3).

[15] The rectangular region of the 2D tape previously visited by the 2D head (the arena) is represented row by row on a 1D tape separated by special markers. The current position of the 2D head

(b) Let $M$ be a deterministic Turing machine that computes $f$ and runs in time $t(n)$. Instead of simulating a universal Turing machine in the block construction, we simulate a Turing machine $M'$ which runs $M$ on input $x$ encoded in the seed frame and acts as program $p_{sb}$ in §5.4. Then the scale of each block is $O(t(n))$, which implies that each block consists of $O(t(n)^2)$ tiles. Now the total number of blocks cannot be more than the running time of $M$ since $M$ outputs every location that corresponds to a block. Thus the total number of tiles is $O(t(n)^3)$.

(c) A similar argument applies to the Wang tiling system as (a) with the following exception. A Wang tiling system can simulate a non-deterministic Turing machine and still be able to output a unique shape. The tiling system can be designed such that if a reject state is reached, the tiling cannot be a valid tiling. For example, the tile representing the reject state can have a bond type that no other tile matches. Thus all valid tilings correspond to accepting computations.

(d) Simulation of Wang tiling systems can, in turn, be done by a non-deterministic Turing machine as follows. Suppose every valid tiling of our Wang tile system has coordinated shape $S$. The simulating NDTM acts similar to the TM simulating self-assembly above, except that every time two or more different tiles can be added in the same location, it non-deterministically chooses one. If the NDTM finds a region containing the seed frame surrounded by *null* bond types, it outputs the shape of the smallest such region and enters an accept state. Otherwise, at some point no compatible tile can be added, and the NDTM enters a reject state. The running time of accepting computations is $O(t(n)^2)$ via the same argument as for (b). □

If, as is widely believed, NDTMs can compute some functions in polynomial time that require exponential time on a TM, then it follows that there exist functions from binary strings to shapes that can be computed much more efficiently by Wang tiling systems than by self-assembly, where efficiency is defined in terms of the size of the coordinated shape produced.

The above relationship between $TIME$ and $TILES$ may not be the tightest possible. As an alternative approach, very small-scale shapes can be created as Wang tilings by using an NDTM that recognizes tuples $(i, j, x)$, rather than one that generates the full shape. This will often yield a compact construction. As a simple example, this approach can be applied to generating circles with radius $x$ at scale $O(n^2)$ where $n = O(\log x)$. It remains an open question how efficiently circles can be generated by self-assembly.

**6.6. Scale Complexity of Shape Languages.** In discussing languages of shapes, it should be possible to discuss complexity in a manner parallel to §6.5. In particular, one may say that a language of shapes is poly-producible by a Turing machine if there is a machine that enumerates it and produces the $n^{th}$ string in time polynomial in $n$, for some ordering of the strings in the language. Similarly, a language of shapes is poly-producible by a tile system if there is some ordering of shapes such that the $n^{th}$ shape is uniquely produced with the total number of tiles used being polynomial in $n$. Analogous definitions can be provided for Wang tiling systems. We conjecture that a strictly smaller class of languages of shapes can be polynomially produced by non-cancerous self-assembly tile systems than by Wang tiling systems.

---

is also represented by a special marker. If the arena is $l \times m$, a single move of the 2D machines which does not escape the current arena requires at most $O(m^2)$ steps, while a move that escapes it in the worst case requires an extra $O(ml^2)$ steps to increase the arena size. We have $m, l = O(t(n))$, and the number of times the arena has to be expanded is at most $O(t(n))$.

**6.7. Other Uses of Programmable Growth.** The programmable block construction is a general way of guiding the large scale growth of the self-assembly process and may have applications beyond those explored so far. For instance, instead of constructing shapes, the block construction can be used to simulate other tile systems in a scaled manner using fewer tile types. It is easy to reprogram it to simulate, using few tile types, a large deterministic $\tau = 1$ tile system for which a short algorithmic description of the tile set exists. We expect a slightly extended version of the block construction can also be used to provide compact tile sets that simulate other $\tau = 2$ tile systems that have short algorithmic descriptions.

To self-assemble a circuit, it may be that the shape of the produced complex is not the correct notion. Rather one may consider finite patterns, where each location in a shape is "colored" (e.g. resistor, transistor, wire, etc.). Further, assemblies that can grow arbitrarily large may be related to infinite patterns. What is the natural way to define the self-assembly complexity of such patterns? Do our results (§4) still hold?

**Acknowledgements.** We thank Len Adleman, members of his group, Ashish Goel, and Paul Rothemund for fruitful discussions and suggestions. We thank Rebecca Schulman and David Zhang for useful and entertaining conversations about descriptional complexity of tile systems.

**Appendix A. Local Determinism Guarantees Unique Production: Proof of Theorem 2.3.**

LEMMA A.1. *If $\vec{A}$ is a locally deterministic assembly sequence of $\mathbf{T}$, then for every assembly sequence $\vec{A'}$ of $\mathbf{T}$ and for every tile $t' = t'_n$ added in $\vec{A'}$ the following conditions hold, where $t = A(pos(t'))$.*

*(i) $inputsides^{\vec{A'}}(t') = inputsides^{\vec{A}}(t)$,*

*(ii) $t' = t$.*

*Proof.* Suppose $t' = t'_n$ is the first tile added that fails to satisfy one of the above conditions. Consider any $D \in inputsides^{\vec{A'}}(t')$. Tile $t_D = A'(D(pos(t')))$ must have been added before $t'$ in $\vec{A'}$ and so $D^{-1} \notin inputsides^{\vec{A'}}(t_D) = inputsides^{\vec{A}}(t_D)$. This implies $D \notin propsides^{\vec{A}}(t)$ and so,

$$(A.1) \quad inputsides^{\vec{A'}}(t') \cap propsides^{\vec{A}}(t) = \emptyset.$$

Now, $\forall D, \Gamma_D^{A'_n}(t') \leq \Gamma_D^A(t')$ because $A'_n$ has no more tiles than $A$ and except at $pos(t)$ they all agree. Equation A.1 implies

$$\Gamma^A_{inputsides^{\vec{A'}}(t')}(t') \leq \Gamma^A_{\mathcal{D}-propsides^{\vec{A}}(t)}(t').$$

Therefore,

$$\Gamma^{A'_n}_{inputsides^{\vec{A'}}(t')}(t') \leq \Gamma^A_{\mathcal{D}-propsides^{\vec{A}}(t)}(t').$$

So by property (2) of definition 2.2, no tile of type $\neq type(t)$ could have been sufficiently bound here by $inputsides^{\vec{A'}}(t')$ and thus $t' = t$. Therefore, $t'$ cannot fail the second condition (ii).

Now, suppose $t'$ fails the first condition (i). Because of property (1) of definition 2.2, this can only happen if $\exists D \in inputsides^{\vec{A'}}(t') - inputsides^{\vec{A}}(t')$. Since $D \notin inputsides^{\vec{A}}(t'))$, $t_D$ must have been added after $t'$ in $\vec{A}$. So since $t_D$ binds $t'$, $D^{-1} \in inputsides^{\vec{A}}(t_D)$ and so $D \in propsides^{\vec{A}}(t)$. But by equation A.1 this is impossible. Thus we conclude $A' \subseteq A$. □

Lemma A.1 directly implies that if there exists a locally deterministic assembly sequence $\vec{A}$ of $\mathbf{T}$ then $\forall A' \in Prod(\mathbf{T}), A' \subseteq A$. Theorem 2.3 immediately follows: If there exists a locally deterministic assembly sequence $\vec{A}$ of $\mathbf{T}$ then $\mathbf{T}$ uniquely produces $A$.

Since local determinism is a property of the *inputsides* classification of tiles in a terminal assembly, lemma A.1 also implies:

COROLLARY A.2. *If there exists a locally deterministic assembly sequence $\vec{A}$ of* **T** *then every assembly sequence ending in A is locally deterministic.*

## Appendix B. Scale-Equivalence and "$\cong$" are Equivalence Relations.

Translation-equivalence is clearly an equivalence relation. Let us write $S_0 \stackrel{tr}{=} S_1$ if the two coordinated shapes are translation equivalent.

LEMMA B.1. *If $S = S_0^d$ and $S_0 = S_m^k$ then $S = S_m^{dk}$.*

*Proof.* $S(i,j) = S_0(\lfloor i/d \rfloor, \lfloor j/d \rfloor) = S_m(\lfloor \lfloor i/d \rfloor/k \rfloor, \lfloor \lfloor j/d \rfloor/k \rfloor) = S_m(\lfloor i/dk \rfloor, \lfloor j/dk \rfloor)$. □

LEMMA B.2. *If $S_0 \stackrel{tr}{=} S_1$ then $S_0^d \stackrel{tr}{=} S_1^d$.*

*Proof.* $S_0^d(i,j) = S_0(\lfloor i/d \rfloor, \lfloor j/d \rfloor) = S_1(\lfloor i/d \rfloor + \Delta i, \lfloor j/d \rfloor + \Delta j) = S_1(\lfloor \frac{i+d\Delta i}{d} \rfloor, \lfloor \frac{j+d\Delta j}{d} \rfloor) = S_1^d(i + d\Delta i, j + d\Delta j)$. □

To show that scale equivalence is an equivalence relation, the only non-trivial property is transitivity. Suppose $S_0^c = S_1^d$ and $S_1^{d'} = S_2^{c'}$ for some $c, c', d, d' \in \mathbb{Z}^+$. $(S_1^d)^{d'} = (S_1^{d'})^d = S_1^{d'd}$ by lemma B.1. Thus, $S_1^{d'd} = (S_0^c)^{d'} = (S_2^{c'})^d$, and by lemma B.1, $S_0^{cd'} = S_2^{c'd}$.

To show that "$\cong$" is an equivalence relation, again only transitivity is non-trivial. Suppose $S_0 \cong S_1$ and $S_1 \cong S_2$. In other words, $S_0^c \stackrel{tr}{=} S_1^d$ and $S_1^{d'} \stackrel{tr}{=} S_2^{c'}$ for some $c, c', d, d' \in \mathbb{Z}^+$. By lemma B.2, $(S_0^c)^{d'} \stackrel{tr}{=} (S_1^d)^{d'}$ and $(S_1^{d'})^d \stackrel{tr}{=} (S_2^{c'})^d$. Then by lemma B.1, $S_0^{cd'} \stackrel{tr}{=} S_1^{d'd}$ and $S_1^{d'd} \stackrel{tr}{=} S_2^{c'd}$ which implies $S_0^{cd'} \stackrel{tr}{=} S_2^{c'd}$ by the transitivity of translation equivalence. In other words, $S_0 \cong S_2$.